\documentclass[manuscript,screen]{acmart}

\AtBeginDocument{%
  \providecommand\BibTeX{{%
    \normalfont B\kern-0.5em{\scshape i\kern-0.25em b}\kern-0.8em\TeX}}}

\usepackage{multirow}
\usepackage{hhline}

\begin{document}

\title{Your signature is your password: Haptic Passwords on Mobile Devices}

\author{Junjie Yan}
\email{junjiey@uw.edu}
\affiliation{%
  \institution{University of Washington}
  \streetaddress{18550 NE 53rd Ct}
  \city{Redmond}
  \state{WA}
  \postcode{98052}}

\author{Tamara Bonaci}
\email{t.bonaci@neu.edu}
\affiliation{%
 \institution{Northeastern University}
 \streetaddress{Khoury College of Computer Sciences, Northeastern University - Seattle 401 Terry Ave N}
 \city{Seattle}
 \state{WA}
 \postcode{98109}}

\author{Howard J. Chizeck}
\email{chizeck@uw.edu}
\affiliation{%
	\institution{University of Washington}
	\streetaddress{Department of Electrical \& Computer Engineering, University of Washington,  Paul Allen Center, 185 E Stevens Way NE AE100R}
	\city{Seattle}
	\state{WA}
	\postcode{98195-0005}}

\renewcommand{\shortauthors}{Yan, et al.}

\begin{abstract}
	In this paper, a new behavioral biometric password based on haptic interaction is described. This "haptic password" system is built on a device that has a force-sensitive screen. The user authentication process leverages maximal overlap discrete wavelet transform (MODWT) based feature extraction along with a customized classifier as well as an adaptive template scheme to accommodate users' variations over time. An experimental study was conducted with 29 subjects and we investigated the authentication performance with multiple user postures under simulated forgery attacks. The performance of the model is evaluated based on authentication accuracies. The results suggest that this proposed haptic password system achieves robust high authentication accuracy and scalability, as well as forgery-proof performance. 
\end{abstract}

\ccsdesc[500]{Security and privacy~Intrusion/anomaly detection and malware mitigation}
\ccsdesc[100]{Security and privacy~Usability in security and privacy}
\ccsdesc[500]{Security and privacy~Biometrics}
\ccsdesc[300]{Human-centered computing~Haptic devices}

\keywords{Authentication, Haptic interfaces, Security}

\maketitle

\section{Introduction}
Existing cyber and cyber-physical systems largely rely on the use of passwords to identify and authenticate human users \cite{wiedenbeck2005passpoints}. Most password-based systems rely on some combination of following concepts to authenticate a user: 1) what you know, 2) what you have and 3) who you are. Current identification and authentication methods can broadly be classified into those that depend on alphanumerical passwords (what you know), and those that use conventional biometric properties of a user (who you are), such as fingerprints, facial recognition, voice data, and iris recognition. Alphanumerical passwords are most widely used since they are easy to implement and the updating process is simple. However, there are drawbacks to the use of alphanumerical passwords. Alphanumerical passwords have a limited possible password space and thus are vulnerable to dictionary and brute-force search attacks  \cite{jablon1997extended}\cite{yan2004password}. Additionally, users often struggle to find a good tradeoff between security and memorizability when dealing with such password systems. More specifically, people tend to: (1) use overly simplistic passwords that are easy to memorize, but also easy to break; (2) reuse their passwords across different systems; (3) not update their passwords regularly\cite{adams1999users}\cite{morris1979password}. On the other hand, biometric passwords release the burden of memorizing passwords from users since they are physical parts of the user. Nonetheless most widely used biometric passwords, such as fingerprint and iris recognition, also have limitations including: (1) potential privacy issues that may arise from the use of these passwords; (2) relatively low accuracy rate; (3) limited ability to update\cite{prevost1999biometrics}. There are also recent concerns about the security of some biometric passwords. In \cite{roy2017masterprint}, the authors suggest that smartphones can easily be fooled by fake fingerprints digitally composed of many common features found in human fingerprints.

These drawbacks motivate the need for new password systems. One possible novel identification and authentication system is based on haptic interaction. In a previous study\cite{yan2015haptic}, we demonstrated that each individual user interacts with a force feedback (haptic) device in a unique way, which can be used as a basis for a new type of biometric identification and authentication, namely, a haptic password. We also showed that the extra force (haptic) information helps generate better authentication performance. Haptic passwords use both `what you know' (graphical information of the password, such as shape and direction) and `who you are' (how the user haptically interacts with the app when entering the password) to authenticate the user. It significantly increases the space of possible passwords, making dictionary and brute force attacks much harder to accomplish. In addition, unlike other biometric-based identification and authentication methods, haptic-based passwords can be updated if the need arises.

In this work, relying on recent technologies, such as force sensitive touchscreens on smartphones and tablets, we built our haptic passwords system on an iPhone 6s. We performed a study with 29 participants of mixed demographics. All subjects enrolled their signatures and self-defined pattern as their passwords. We then developed a maximal overlap discrete wavelet transform (MODWT)\cite{whitcher2000wavelet} based feature extraction and combined it with a customized classification strategy to fulfill the authentication task. We also developed an adaptive template update scheme to accommodate user's variation over time. We tested our authentication technique and collected detailed data on the authentication accuracy and how different writing postures (i.e. hand and device position), and day-to-day variation affects the system performance. We simulated skilled forgery attacks and tested the proposed system's vulnerability to such attacks. 

In summary, the main contributions of this work are:

\begin{enumerate}
	
	\item Development a MODWT based feature extraction and customized classifier to realize the user identification and authentication.
	\item Development and demonstration of an app on iPhone 6s to implement the haptic password system, and an experimental evaluation with 29 participants.
	\item Development of an adaptive template update scheme to account for variations of a user's password over time.
	\item Experimental demonstration that the proposed haptic password system is able to achieve high authentication accuracy and scalability as well as forgery-resistance.
\end{enumerate}


\section{Related Work}
Bonneau et al. \cite{bonneau2012quest} conducted a comprehensive survey that evaluated two decades of proposals to replace text passwords for general-purpose user authentication with a set of twenty-five usability, deployability and security benefits that an ideal password scheme should provide. Their results lead to key insights about the difficulty of replacing the existing text passwords. The report that not only does no known scheme come close to providing all desired benefits, but none retains the full set of benefits that text passwords already provide. 

Chiasson et al.\cite{chiasson2007graphical}, Jablon et al.\cite{jablon1997extended} and Wiedenbeck et al.\cite{wiedenbeck2005passpoints} proposed using graphical based password systems to authenticate users. 
The main idea of these systems is to let the user click on a few chosen regions of an image and, based on the clicked location, to authenticate the user. Such systems provide benefits over alphanumeric and biometric passwords. In general, users are able to better memorize graphical passwords. Additionally, the possible graphical passwords space is much larger than that of alphanumeric passwords. One major concern, however, is that graphical passwords are still vulnerable to forgery attacks (i.e. shoulder surfing attacks) \cite{lashkari2009shoulder}.

To overcome this issue and increase the possible password space, several studies have been done to authenticate the user by implementing haptic-based information. Bianchi et al. \cite{bianchi2011spinlock}\cite{bianchi2012counting} proposed authentication schemes that combine both visual and haptic/audio cues, such as vibration or specific sound effect, as password input. There are two major limitations in their work. They rely on extra hardware, such as a tactile wheel or earphones, to generate cues. Moreover, the authentication speed is relatively low (more than 10 seconds). Krombholz et al. \cite{krombholz2016use} developed a force-PINs technology that enhances traditional 4-digits or 6-digits PINs with tactile features using pressure-sensitive touchscreens as found in modern consumer hardware. Besides 0-9 for each digit, they enrich the password information with force pressure. While their approach offers an "invisible" force layer to the password, the possible password space is limited as the force is discretized into "deep" and "shallow" press only. If an attacker learns the PIN, such as through shoulder-surfing, the password can be hacked through brute-forcing. Additionally, the haptic channel introduced in both aforementioned work adds extra memorization burden on the user. 

Another way to authenticate the user is through their signature\cite{faundez2007line}\cite{huang2003stability}\cite{nanni2010combining}\cite{ohishi2001pen}. With the development of force-sensing technology, not only the position information about the signature but also the `hidden' haptic (force) information can be captured. This extra dimension of data enhances the security of these systems. Currently, most signature verification algorithms are based on dynamic time warping (DTW) to realize user identification and authentication\cite{impedovo2008automatic}. For an input signal and a template of different lengths, the DTW algorithm is able to find the best matching path, in terms of the least global distance, based on dynamic programming (DP)\cite{sankoff1983time}. However, DTW has several drawbacks when applied in online signature verification. First, DTW is a computationally expensive algorithm\cite{feng2003online}. This is because the DTW performs non-linear warping on the entire signal. The computational cost is proportional to the square of the signal length\cite{sankoff1983time}. Second, the DTW algorithm is required to store the raw signature signal as a template to fulfill the matching process for a test signature signal. In case the malicious party hacks the database, the malicious party will be able to obtain complete information of each user's signature. Aware of this issue, in \cite{iglesias2008characterizing} \cite{orozco2008experiments}\cite{orozco2006haptic}\cite{alsulaiman2008user} \cite{sae2014online}, global statistical properties such as mean, variance, correlation of the recorded signature data\cite{iglesias2008characterizing} \cite{orozco2008experiments}\cite{orozco2006haptic}\cite{alsulaiman2008user} and histogram of trajectory direction and location \cite{sae2014online} were used as password features to classify the user. The drawbacks of the aforementioned techniques are the loss of signature transient properties in the feature extraction process.

In \cite{frank2013touchalytics},  a continuous authentication based on how a user interacts with the touch screen is proposed. The author investigated the user's touch actions during navigation maneuvers to continuously authenticate the user. The major limitation of this work is that it requires a relatively long enrollment period and the experimental findings of this work disqualify this method as a standalone authentication mechanism for long-term authentication.

In \cite{afsar2005wavelet} \cite{fahmy2010online} and our previous work \cite{yan2015haptic}, the discrete wavelet transform (DWT) is implemented to extract features from the user's signature in order to realize user authentication. The key benefit of the DWT is that it captures both frequency and localized (in time) information. However, there is a constraint for DWT that the length of the signal has to be a multiple of a power of two and hence resampling or zero-padding the original signature signal is required. These modifications alter the original properties of the given signature signal and will potentially affect the authentication performance.

Additionally, in most aforementioned works, conventional classification algorithms, such as neural network, support vector machine or random forest, are implemented. One major limitation of these classification algorithms in signature verification applications is that in the training phase they heavily rely on the choices of both positive and negative samples. Although the choices of positive samples are straightforward as we can use the genuine signatures, \emph{how to choose negative samples to train the classifier is a tricky problem}. First, using forgeries (as in \cite{justino2005comparison}) as negative samples to train the classifier is not applicable in the real-life scenario, since the forgeries require extra effort to generate. On the other hand, although using different subjects' signatures or predefined patterns as negative samples is feasible, this may lead to the classifier under-estimate the difficulty of the classification problem, since different signatures can easily be classified graphically while it is more challenging to classify genuine signature and its forgery. Another limitation of conventional classifiers is a lack of adaptivity. Galbally et al. \cite{galbally2013aging} demonstrated that the dynamic property of human's signature may vary over time. Therefore, to achieve reliable long-term use of the signature-based password system, it is crucial to develop a mechanism to account for the variation. However, for these conventional classifiers, a new training set has to be reconstructed to address the signature variation over time and the entire training process for the classifier needs to be performed again.

Recently, Harbach et al. \cite{harbach2016anatomy} conduct a detailed real-world study of the smartphone unlocking and provide a benchmark of current smartphone authentication mechanisms. In their study, it is shown that on average, participants unlock their phones more than 40 times per day. Also according to the participants' subjective survey, most participants put more emphasis on the authentication speed of the unlock system. These results indicate that in the real-world application, the time needed to fulfill the authentication process is a critical factor regarding the usability of the password system.

\section{Problem Statement and Goals}

\subsection{Problems}

Problem 1: Forgery attacks.\\

\begin{figure}[thpb]
	\centering
	\includegraphics[width=1\columnwidth]{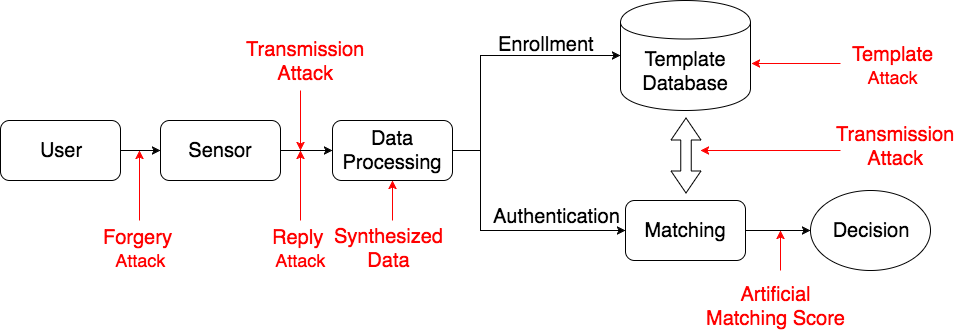}
	\caption{Attacks on biometric authentication systems}~\label{fig:prob_figure1}
\end{figure}

The general structure of a biometric authentication system is depicted in Figure \ref{fig:prob_figure1}\cite{xiao2007technology}. Various types of attacks can be launched to compromise the system at different stages. 

In this paper, we focus on demonstrating that our proposed haptic password system is resistant to various forgery attacks. Haptic information, such as force and velocity, is able to classify and authenticate the user and prevent the forgery attack. Our proposed feature extraction technique can mitigate template attacks. As discussed later, the feature extraction process is non-invertible, so even if the template is hacked, a malicious party still has no information about the genuine password.\\


Problem 2: The lack of negative samples to train the classifier.\\

During the classifier training phase, only positive samples (genuine password from the user) are available, thus conventional classifiers, whose training procedures depend on both positive and negative samples (forgeries), are not appropriate for this application. Customized classifiers need to be developed to classify different users while achieving forgery attack resistance. \\

Problem 3: Randomness of Haptic Interaction.\\

The randomness of a human during haptic interaction also introduces extra challenges to the authentication system. First, during the enrollment phase for each user, the system must be trained based on the training set of the given user and generate a template database. However, there potentially will be `bad' or `inconsistent' training sets in the enrollment phase. Moreover, the dynamic properties of how a human haptically interacts with the device will vary over time \cite{galbally2013aging}. It is crucial to develop a method to adaptively update the template database to accommodate the variation over time so that the authentication system does not become stale.

\subsection{Goals}
The main goals of our study are to analyze the classification accuracy and forgery resistance of the proposed password system, and its robustness over time. More specifically, we want to answer the following questions:

1) What is the probability that a genuine password gets rejected?

2) What is the probability that a forged password gets accepted?

3) How do a user's postures and forms of passwords affect the authentication performance?

4) How robust is the authentication system over time?

\section{Haptic Passwords System}
\begin{figure}[thpb]
	\centering
	\includegraphics[width=1\columnwidth]{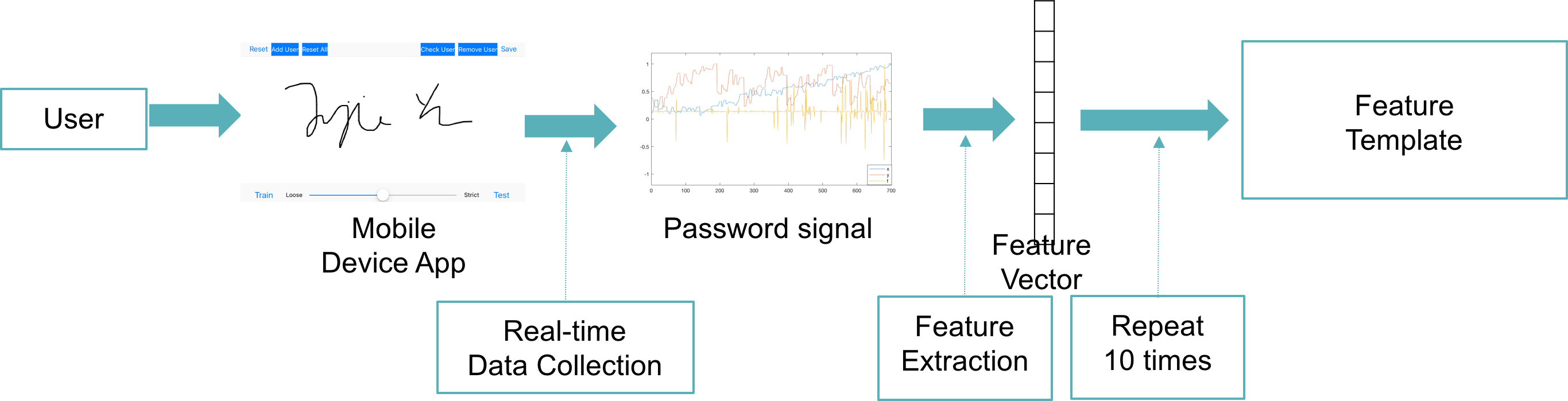}
	\caption{Haptic Password System Training Process. The users first interact with the mobile device app to enroll their passwords. The position and force data of the passwords are collected in real-time. We then conducted a MODWT based feature extraction to obtain the feature vector. The whole process is repeated 10 times in order to generate the feature template set for each user.}~\label{fig:hapsys_figure2}
\end{figure}

\begin{figure}[thpb]
	\centering
	\includegraphics[width=1\columnwidth]{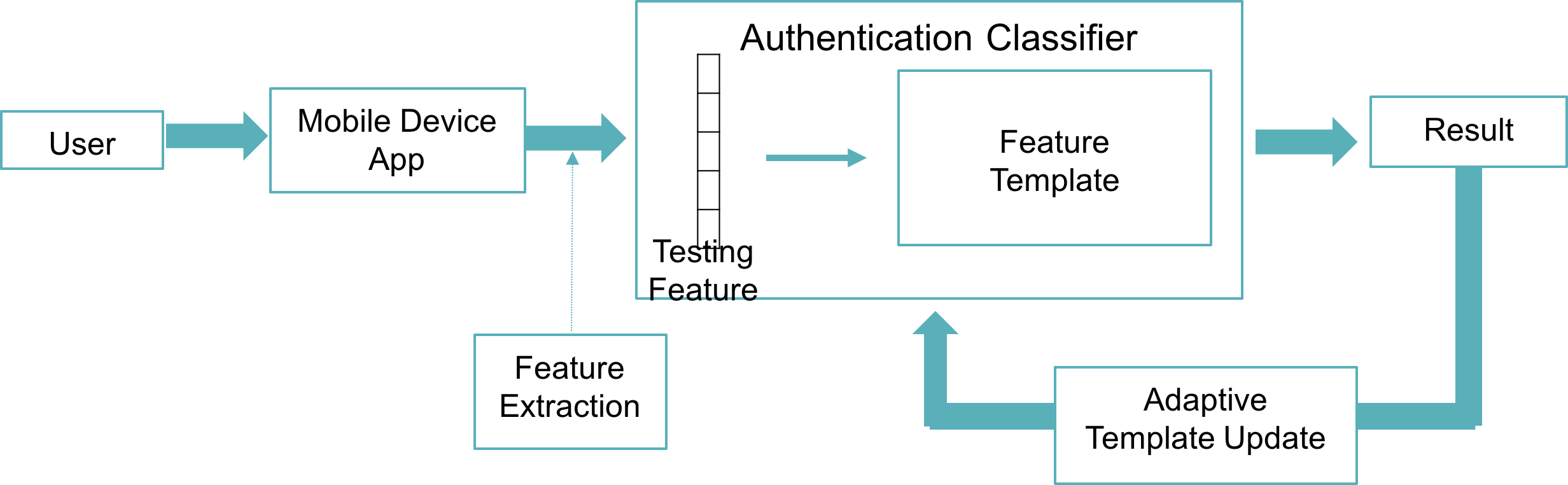}
	\caption{Haptic Password Authentication Process. The users enter their passwords on the mobile device app. The position and force data are collected and the corresponding feature vectors are obtained. We then generate the authentication result based on the analysis between the feature vector of the testing password and the feature template set. Once the testing password passes the authentication, the feature template is adaptively updated.}~\label{fig:hapsys_figure3}
\end{figure}

In order to achieve the goals and answer these questions, we developed a haptic password system as shown in Fig. \ref{fig:hapsys_figure2} (training process) and Fig \ref{fig:hapsys_figure3} (authentication process). This haptic password system consists of three main parts: 1) real-time data collection, 2) feature extraction, and 3) user authentication. In the training process, a user will enter the password (i.e. signature) 10 times in order to train the authentication classifier. The mobile app will collect position and force data in real-time and extract corresponding features. A classifier will be trained based on the extracted features. In the authentication process, we then use the trained classifier to fulfill user authentication. We also developed an adaptive template update scheme to accommodate a user's password variation overtime.

\section{Experiment and Data Collection}
\begin{figure}[thpb]
	\centering
	\includegraphics[width=0.48\columnwidth]{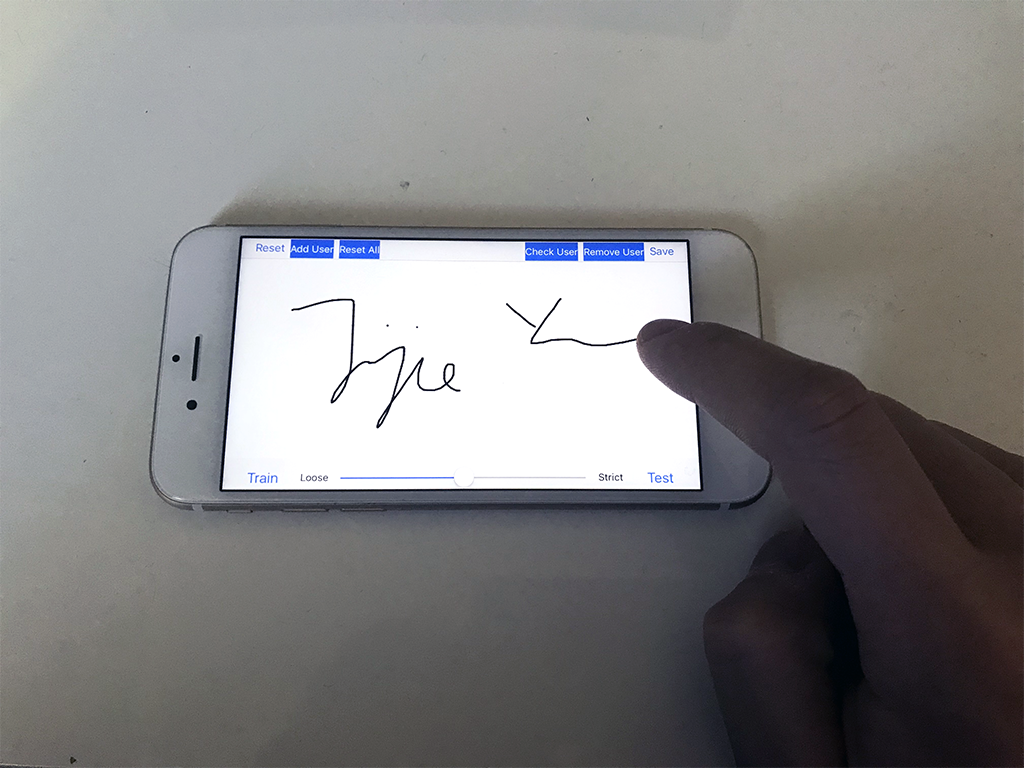}
	\includegraphics[width=0.48\columnwidth]{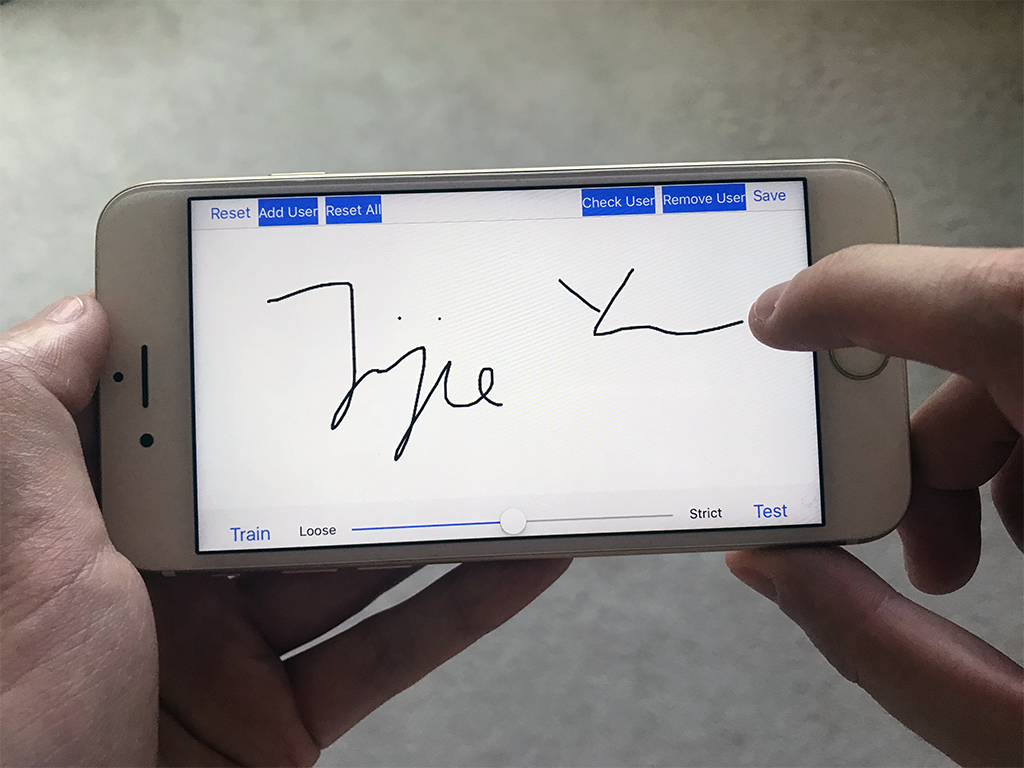}
	
	\caption{Experiment: User Sign Signature in the Static Posture (Left) and the Holding Posture (Right)}~\label{fig:exp_figure1}
\end{figure}

\begin{figure}[thpb]
	\centering
	\includegraphics[width=0.48\columnwidth]{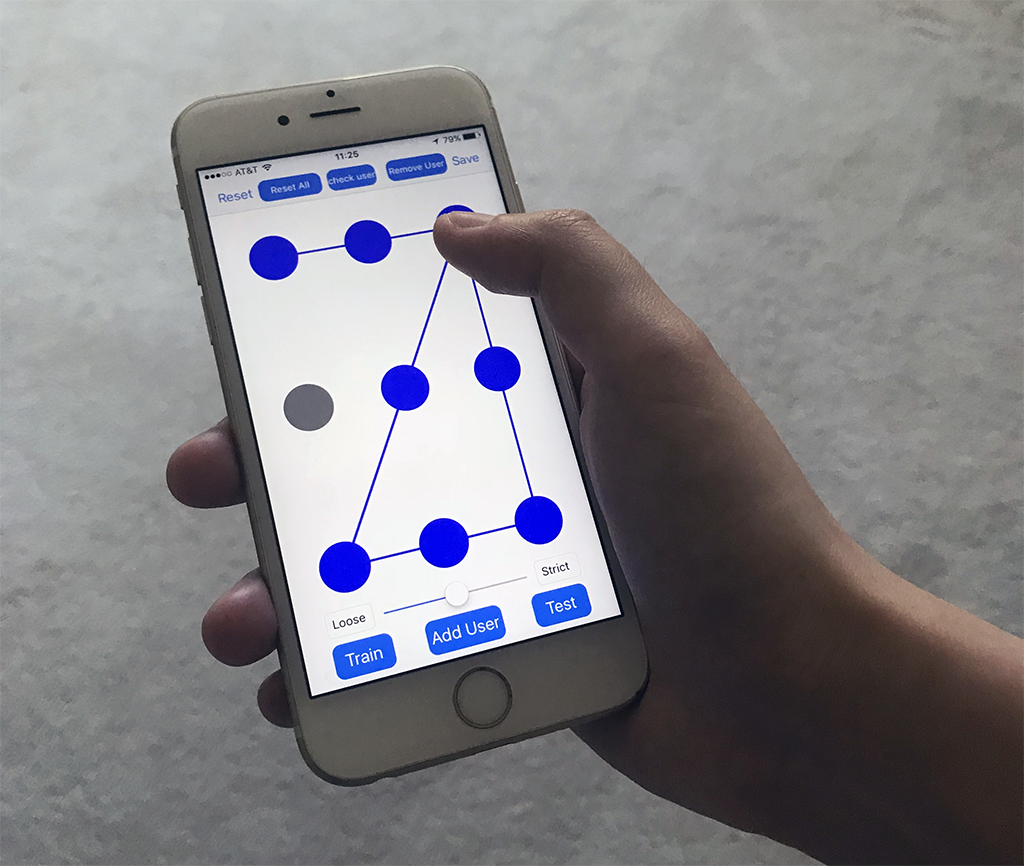}

	\caption{Experiment: User Enter Pattern Password while Holding the Phone }~\label{fig:exp_figure1.1}
\end{figure}

We conducted human subjects experiments on an iPhone 6s where subjects used their signature as passwords by using their finger to sign on an iPhone (as shown in Fig. \ref{fig:exp_figure1}) or drawn a subject-defined pattern (as shown in Fig. \ref{fig:exp_figure1.1}). We also simulated a forgery attack where each subject was asked to forge the passwords of four other subjects. The video of how the other subject's generated the password was available to the forger. We then tested the system's resistance to these forgery attacks. All experiments were conducted with the approval of the UW Institutional Review Board.

\subsection{Experiment Setup}

There are two sections in this experiment. In the first section, subjects used their signature as a password, while in the second section, subjects defined their own pattern as a password. 

First, subjects were asked to explore the app environment and practice their passwords for 5 to 10 times in order to get used to the interface, reduce the effects of learning and to generate consistent passwords. 

Each subject was asked to fulfill the following tasks:\\
1-1) Sign the signature with the phone being on a static flat surface; (referred as \textbf{signature with static device} in the following)\\
1-2) Forge other subjects' signatures with the phone being on a static flat surface; \\
1-3) Sign the signature with the phone being held in hand; (referred as \textbf{signature with holding device} in the following) \\
1-4) Forge other subjects' signatures with the phone being held in the hand;\\

Task 1-1) and 1-3) aim to test how different postures affect the authentication results and task 1-2) and 1-4) simulate a forgery attack and test system resistance to this type of attack. The detailed attack model will be discussed in the following section.

Similarly, in the second section, each subject was asked to fulfill the following tasks:\\
2-1) Enter the pattern with the phone being held in the hand; (referred as \textbf{pattern with holding device} in the following) \\
2-2) Forge other subjects' pattern with the phone being held in the hand;\\

For the pattern password entry, the case of having the phone placed on a flat surface was not considered because most users tend to hold the phone in one hand and use their thumb to finish the pattern (as shown in Fig. \ref{fig:exp_figure1.1} ). This is not possible if the phone is on a flat surface.

For each subject, the experiment is completed in two consecutive days to examine the system authentication performance over time. On day 1, the subject completed all tasks listed above, while on day 2, only task 1-1), 1-3) and 2-1) were completed. 

\subsection{Attack Model}
The goal of attacks against the authentication system is to impersonate a specific user. In this work, we focus on the forgery attack where a malicious individual pretends to be someone else by providing a forged password of a legitimate user. 

\textbf{Forgery Attack}: The malicious individual is able to watch how the victim enters the password and generates forgeries based on the observation.

In order to simulate such an attack, all genuine passwords are video recorded in real-time. Each subject (simulated `malicious individual') is able to watch videos of previous subjects entering their password and forge those passwords based on this observation. In this way, the simulated `malicious individual' has access to not only the static graphical but also the dynamic entry of each password.



.

\subsection{Data Collection}

In this experiment, we use an iPhone 6s to collect user haptic password data. The phone's force-sensitive touch technology allows us to collect not only the contact position information but also how much force is applied as the users enter their haptic password.

We collect the following data in real time:\\
1) Position (in pixel) of contact on screen $(x,y)$;\\
2) Applied forces $(f)$;\\

All data is recorded at 60 Hz. The iPhone 6s app starts recording data when the user makes contact with the screen and stops when the user presses the `save' button. Each password signal data is then truncated to the point where the last contact happens.

In summary, for each subject, the experimental data collected includes:\\
1) 30 genuine signatures with the device on the flat surface (Day 1: 20 sets, Day 2: 10 sets)\\
2) 30 genuine signatures with the device held in hand (Day 1: 20 sets, Day 2: 10 sets)\\
3) 30 genuine pattern with the device held in hand (Day 1: 20 sets, Day 2: 10 sets)\\
4) 20 forgeries of signatures with the device on the flat surface (from 4 different subjects, 5 forgeries per subject)\\
5) 20 forgeries of signatures with the device held in hand (from 4 different subjects, 5 forgeries per subject)\\
6) 20 forgeries of pattern with the device held in hand (from 4 different subjects, 5 forgeries per subject)\\

For the simulated forgery attacks, each subject was asked to watch videos of the previous 4 subjects' genuine signatures. We then allowed the subjects to replay the videos and practice forgery until they were satisfied with their result. After that, they started to generate corresponding forgeries for each victim subject. 5 forgeries are generated for each victim subject in each posture(i.e. $i^{th}$ subject will forge $(i-4)^{th}\sim(i-1)^{th}$ subject's signatures). The first 4 subjects were then asked to participate in follow-up experiments where they forged the last 4 subjects' signatures (i.e. the first subject will forge $(N-3)^{th}\sim(N)^{th}$ subject's signatures, where N is the total number of subjects). In this way, each subject's genuine signatures in each posture were forged 20 times by 4 different subjects.

\subsection{Subjects Demographics}
\begin{figure}[thpb]
	\centering
	\includegraphics[width=0.48\columnwidth]{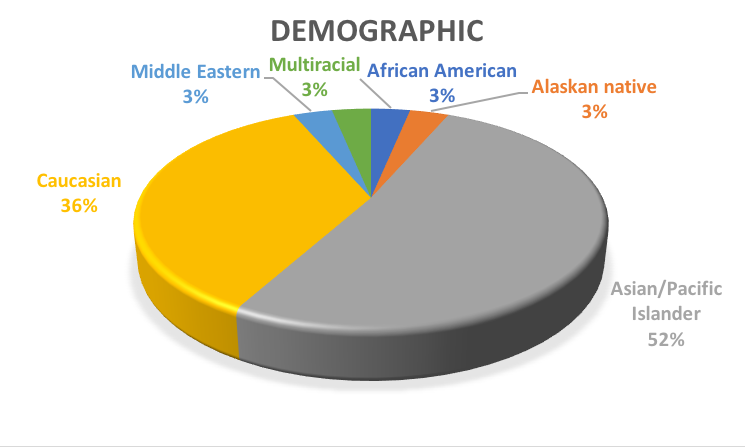}
	\includegraphics[width=0.48\columnwidth]{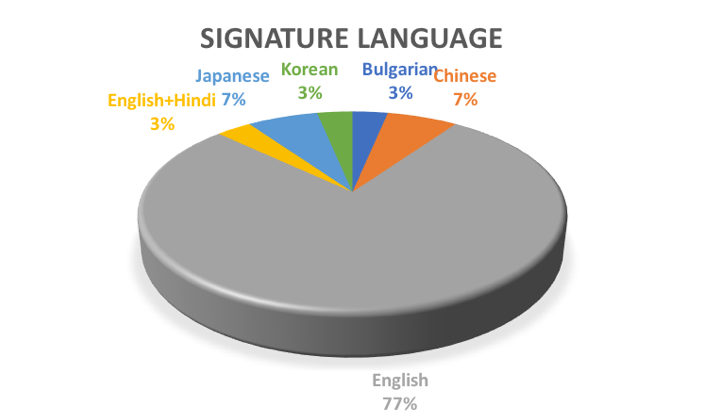}
	\caption{Subjects Demographic (Left) and Signature Language (Right) Distribution}~\label{fig:exp_figure2}
\end{figure}

Our analysis is based on data collected from experiments involving 29 participants. All of the individuals did not have any physical impairment that affected their ability to sign their names. The detailed demographics of all subjects is shown in Table \ref{tab:exp_table1}. The subject racial and signature language distribution are shown in Fig. \ref{fig:exp_figure2}.

\begin{table}[h]
	\begin{center}
		\caption{Subjects Demographics}
		\label{tab:exp_table1}
		\scalebox{1}{
			\begin{tabular}{|c |c|}
				\hline
				Sample Size & 29  \\
				\hline
				Sex & 14 Females; 15 Males  \\
				\hline
				Age Range & 18 to 62 \\
				\hline
				Age (Mean $\pm$ SD) & 26.5 $\pm$ 8.7\\
				\hline
				Handedness & 4 Left; 25 Right\\
				\hline
				
			\end{tabular}
		}
	\end{center}

\end{table}	

\section{Feature Extraction}
In order to accomplish user authentication given the haptic passwords, the first step is to extract features. The choice of elements in the feature vector significantly affects the performance of the classifier. In our previous work \cite{yan2015haptic}, we implemented the discrete wavelet transform (DWT) to fulfill the feature extraction of the signature signal. The key benefit of the DWT is that it captures both frequency and localized (in time) information. However, for DWT the length of the signal has to be multiples of a power of two. If we resample the password signal to meet this requirement, this alters the original property of the given password signal and potentially increases the similarity between genuine passwords and forgeries (since the original length of genuine passwords and forgeries is most likely different, while after resampling, they will be the same). To avoid this problem, in this work, we use the maximal overlap discrete wavelet transform (MODWT)\cite{percival2006wavelet}, a modified discrete wavelet transform, to deal with the signature signals. Unlike the DWT, MODWT is defined naturally for all signal lengths. \emph{This allows us to perform feature extraction on the original signature signal without resampling.}\\

Additional security of the stored feature template is obtained because this feature extraction process is non-invertible. Even if the malicious party is able to hack the template database, it is impossible to reconstruct the original password from the features. 

\subsection{State vector}
In addition to the position of contact on the screen $(x,y)$ and applied forces $(f)$ we obtained from the experiment, we calculate additional quantities to formulate the state vector as follows:

1) Signature trajectory velocity
$$ 
v_x(t)=x(t)-x(t-1), v_x(0)=0 \eqno{(1)}
$$
$$
v_y(t)=y(t)-y(t-1), v_y(0)=0 \eqno{(2)}
$$

2) Motion Acceleration
$$
a(t)=\sqrt{v_x^2(t)+v_y^2(t)}-\sqrt{v_x^2(t-1)+v_y^2(t-1)}, a(0)=0\eqno{(3)}
$$

3) Trajectory direction
$$
\theta(t)=atan(\frac{v_y(t)}{v_x(t)}), \theta(0)=0 \eqno{(4)}
$$

4) Trajectory direction change
$$
\omega(t)=\theta(t)-\theta(t-1),\omega(0)=0 \eqno{(5)}
$$

The state vector is then constructed as $s=(x,y,f,v_x,v_y,a,\theta,\omega)$.

\subsection{Maximal Overlap Discrete Wavelet Transform}
In this section, we briefly review the general concept of MODWT. Similar to the DWT, the MODWT coefficient of signal $X$ is calculated by passing it through a series of filters generated from a mother wavelet filter. The mother wavelet filter $g$ is a low-pass filter that satisfies the standard quadrature mirror condition \cite{percival2006wavelet}\cite{shensa1992discrete}

$$
G(z)G(z^{-1})+G(-z)G(-z^{-1})=1\eqno{(6)}
$$

where $G(z)$ denotes the z-transform of the filter $g$. Its complementary high-pass filter can be obtained as

$$
H(z)=zG(-z^{-1})\eqno{(7)}
$$

These mother wavelet filters are then used to generate the series of scaling filters $H$ and wavelet filter $G$
$$
H_{i+1}(z)=H(z^{2^i})G_i(z)\eqno{(8)}
$$
$$
G_{i+1}(z)=G(z^{2^i})G_i(z)\eqno{(9)}
$$

with initial condition $G_0(z)=1$.

Let $h_j(k)$ and $g_j(k)$ be the time-domain representation of $H_j(z)$ and $G_j(z)$ respectively. The MODWT scaling filter $\widetilde{h}_{j}(k)$ and wavelet filter $\widetilde{g}_{j}(k)$ are:

$$
\widetilde{h}_{j}(k)=\frac{h_j(k)}{\sqrt{2}}\eqno{(10)}
$$
$$
\widetilde{g}_{j}(k)=\frac{g_j(k)}{\sqrt{2}}\eqno{(11)}
$$

Let $X$ be a time series of length $N$. Similar to standard DWT, the MODWT scaling coefficient $\widetilde{V}_j(k)$ and wavelet coefficient $\widetilde{W}_j(k)$ at the $j^{th}$ level is calculated as shown in (12) and (13)

$$
\widetilde{W}_j(k)=\sum_{l=0}^{L_j-1}\widetilde{h}_j(l)X((k-l)\;\;mod\;\; N)\eqno{(12)}
$$
$$
\widetilde{V}_j(k)=\sum_{l=0}^{L_j-1}\widetilde{g}_j(l)X((k-l)\;\;mod\;\;N)\eqno{(13)}
$$

\begin{figure}[thpb]
	\centering
	\includegraphics[width=1\columnwidth]{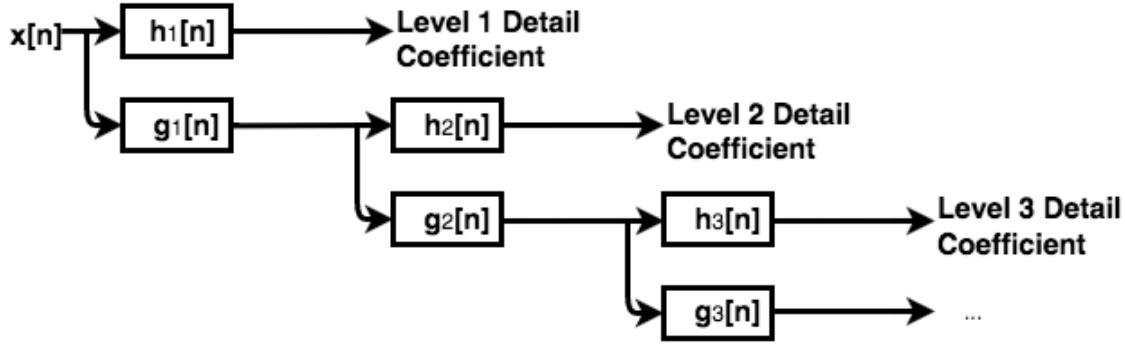}
	\caption{MODWT sub-band decomposition}~\label{fig:modwt_figure1}
\end{figure}

Here $L_j$ is the width of $j^{th}$ level filter. Here $L_j=(2^j-1)(L-1)+1$, and L is the width of the initial mother wavelet filter. Figure \ref{fig:modwt_figure1} shows the block diagram of the MODWT process.

The input signal $X$ is decomposed into high and low frequency components at each level and they are regarded as detail coefficients and approximate coefficients of that corresponding level respectively. \\

\subsection{Feature Vector}
The feature vector for each signature signal is obtained in the following steps:

1. The MODWT is applied to each channel (position, velocity, the force applied, acceleration, direction and direction change) in the state vector $s$ separately. For each channel, 5 levels of decomposition are accomplished and we obtain 5 levels of wavelet coefficients $W_1 \sim W_5$ and level 5 scaling coefficients $V_5$.

2. The following statistical features are used to represent the time-frequency distribution of the signature signals:
\begin{itemize}
	
	\item Mean of the absolute values of the coefficients in each sub-band.
	\item Standard deviation of the absolute values of the coefficients in each sub-band.
	\item Maximum of the absolute values of the coefficients in each sub-band.
	\item Ratio of the absolute mean values of adjacent sub-bands.
	
\end{itemize}

More specifically, for $i=1,2,3,4$,we define
\begingroup\makeatletter\def\f@size{8}\check@mathfonts
$$ 
v_i=[mean(|W_i|),std(|W_i|),max(|W_i|),mean(|W_i|)/mean(|W_{i+1}|)] \eqno{(14)} \
$$
\endgroup

for $i=5,6$, we define
\begingroup\makeatletter\def\f@size{8}\check@mathfonts
$$
v_5=[mean(|W_5|),std(|W_5|),max(|W_5|),mean(|W_5|)/mean(|V_{5}|)] \eqno{(15)}\\
$$
$$
v_6=[mean(|V_5|),std(|V_5|),max(V_5)], \eqno{(16)}
$$
\endgroup

Therefore, the feature vector of each channel in signature signal $s$ is $f_i=[v_1,v_2,v_3,v_4,v_5,v_6]$. The length of $f_i$ is 23. Then the complete feature vector of each signature signal is obtained by concatenating all 8 channels together as $F=[f_1,f_2,...,f_8]$. The length of F is $23\times8=184$.

\section{Training, Testing and Evaluation}
Given the experiment dataset, we aim to answer the following questions:

1) How does a user's password day-to-day variation affect the authentication performance?

2) Can we achieve similar authentication performance when the user enters the password in different postures?

3) How resistant is the authentication to forgery attacks?\\

In order to answer the question above, we divide the dataset into a training set and a testing set. \\

The first 10 signatures with static/holding device and subject-defined patterns in day 1 are used to train the corresponding classifiers and tune their parameters. The remaining 10 signatures with static/holding device and subject-defined patterns on day 1 are used as the same-day testing set while 10 signatures with static/holding device and subject-defined patterns on day 2 are used as the second-day testing set. This allows us to explore the effect of passwords day-to-day variation on the authentication accuracy.

\section{User Authentication}
Recall that in order to realize user authentication, we need to deal with the following three problems: 1) the lack of negative samples to train the classifier; 2) the randomness of the human, and 3) the forgery attacks. Therefore, in this work, we develop an authentication classifier without using negative samples. The authentication result is based on the distance between the testing feature vector and the feature vectors in the template set. The template set is generated in the training process. The authentication processes based on two distance metrics, Euclidean distance and Hamming distance, are developed and tested. We also proposed an adaptive template update mechanism to address the issue of the human randomness over time.

\subsection{Euclidean Distance}
In this section, we calculate the distance between the testing password and the password template based on the Euclidean distance.

A given user's training password feature set is normalized, and the mean and standard deviation of the training feature set is stored to normalize the testing data. Because the dimensionality of the feature vector is large (23 per channel), simply treating each dimension equally for all different users is not appropriate. Based on our observation, for a given user, some feature dimensions vary more significantly than others across different feature vectors. Therefore, we formulate the following optimization problem to find the weighting for each dimension such that for a given user, the password features in the training set are weighted in such a way that makes the feature vectors most consistent to his/her/theirself.

First, given vector $X=[x_1,x_2,\dots,x_n], Y=[y_1,y_2\dots,y_n]$, we define operation $\otimes$ as $X\otimes Y=[x_1y_1,x_2y_2,\dots,x_ny_n]$. Let the weighting parameter for the $s^{th}$ user be $w_s=[w_{s,1},w_{s,2},\dots,w_{s,n}]$

\begin{equation*}
	\begin{aligned}
		& \underset{w_s}{\text{minimize}}
		& & \sum_{i=1}^{N} \sum_{j=i+1}^{N} d(w_s\otimes F_{s,i},w_s\otimes F_{s,j}) \\
		& \text{subject to}
		& & w_{s,i}\ge0, \; i = 1, \ldots, n.\\
		&&& \sum_{i=1}^{n}w_{s,i}=1
	\end{aligned}
\end{equation*}

where $N$ is the number of samples in the training set, $n$ is the dimensionality of the feature vector and $F_{s,i}$ is the $i^{th}$ feature vector in the training set of the $s^{th}$ user. $d(X,Y)$ calculates the Euclidean distance between vector $X$ and $Y$.

For the password authentication, we use the passwords in the training set as templates and match the testing password to the templates in order to obtain the matching distance $d_E$. For the user $s$ authentication, given the genuine password template set $\{F_{s,i}\}$ and testing password $F_{test}$, we are able to obtain $d_i=d(w_s\otimes F_{test},w_s\otimes F_{s,i})$  and $D_E=[d_1,d_2,...,d_N]$. The distance set $D_E$ contains the Euclidean distance between the testing password and each template password. We can then generate the matching distance based on the distance set $D_E$.

However, due to the random changes in human signatures, simply calculating the matching distance by summing all the distance together may not generate a reliable result. The major reasons are two-fold.
\textbf{1) `Bad' samples can be generated during the training process.} 
We observed that users sometimes enter poorly in the training set and thus the corresponding sample is inconsistent with the genuine ones\footnote{These bad samples were entered, in some cases, to deliberately confound the method. This is a problem with knowledgeable test subjects.}. 
\textbf{2) Multiple ways of generating the sample.}
Besides 'bad' samples in the training set, we also noticed that some subjects have multiple ways to generate their passwords and each type of passwords is self-consistent. For example, as for the letter "t" in signature password, the same subject has two different stroke sequences: a) horizontal line followed by the "$l$" shaped stroke; b) the "$l$" shaped stroke followed by a horizontal line. 

When calculating the matching distance, if we use the entire training set as templates, those poorly entered samples and different types of samples will potentially increase the matching distance of a genuine testing password and degrade the authentication performance. 

Although `bad' samples can potentially be detected as outliers and removed from the template set to reduce the intra-template variation, this fix does not work for the case of multiple ways of generating passwords, as all types of passwords should be regarded as genuine. 

Therefore, instead of manipulating the training template, we define the matching distance as follows. For any vector $S\in\mathbb{R}^N$ and $k<N$, we define operation $SumMin(S,k)$ as to sum the $k$ smallest elements in vector $S$. The Euclidean based matching distance $d_E$ of the given testing password is calculated as in (17)

$$
d_E=SumMin(D_E,k) \eqno{(17)}
$$
where $k\in [1,2,...,N]$ is a tuning parameter. Smaller $d_E$ means that the testing password is more similar to the genuine ones in the template set. The testing password will be authenticated when the matching distance is below the predefined threshold $T_{D_E}$.

With the tuning parameter $k$, testing passwords will be matched to those `good' templates or templates with the same type of password. By doing so, we will be able to mitigate the effect of the human's randomness in the training phase.

\subsection{Hamming Distance}
Next we consider calculating the matching distance using the Hamming Distance between the testing password and the password template.

Recall that the Hamming distance between two vectors of equal length is defined as the number of positions at which the corresponding symbols are different.
Therefore, in order to use Hamming distance, the definition of "different" between features in the testing password feature vector and the password template needs to be defined.

First, the mean and standard deviation of the given user's training password feature set is obtained as $\mu = [\mu_1,\mu_2,...,\mu_n]$ and $\sigma = [\sigma_1,\sigma_2,...\sigma_n]$, where $n$ is the dimensionality of the feature vector. We then update the mean and standard deviation in the following ways in order to mitigate the effect of the `Bad' training password as mentioned in the previous section. We assume that features from those `Bad' training passwords are outliers. Therefore, for each feature obtained from the $j^{th}$ training password $f_{i,j}$,  if $|f_{i,j} - \mu_i|\ge 2\sigma_i$, the corresponding feature is discarded. The updated mean $\hat{\mu}$ and standard deviation $\hat{\sigma}$ is calculated by using the remaining features.

The testing password feature vector $F_{test} = [f_1,f_2,...,f_n]$ is normalized based on $\hat{\mu}$ and $\hat{\delta}$ as in (18)

$$
\tilde{f_i}=\frac{f_i-\hat{\mu_i}}{\hat{\sigma_i}}
$$
$$
\tilde{F}_{test} = [\tilde{f_1},\tilde{f_1},...,\tilde{f_n}]\eqno{(18)}
$$

If $|\tilde{f_i}| > T_f$, the corresponding feature in the testing password is regarded as different from the password template, where $T_f$ is a pre-defined threshold for feature difference. The Hamming distance $d_H$ between a feature vector of a testing password and the password template set is defined as the number of features that can be regarded as different. The testing password will be authenticated when the Hamming distance $d_H$ is below the predefined distance threshold $T_{D_H}$.

\subsection{Adaptive template update}

Another complication is the variation of a user's password over time (due to age, changes in muscle strength or for any other reason). In order to address this issue, we developed two adaptive template update schemes for the Euclidean distance and Hamming distance based authentication process.\\

1. Euclidean Distance

The adaptive template update scheme for the Euclidean distance based authentication consists of the following steps:

1) Calculate the Euclidean based matching distance $d_E$ between the given testing password and the password templates as in $(17)$.

2) If the matching distance is greater than the threshold $T_{D_E}$, reject the given testing password.

3) If the matching distance is smaller than the threshold $T_{D_E}$, authenticate the given testing password. Meanwhile, find the password template in the template set that has the largest distance to the current testing password, and replace it with the current new one.

In this way, we will be able to adaptively update the signature template set to accommodate the user's password variation over time.\\

2. Hamming Distance

The adaptive template update scheme for the Hamming distance based authentication consists of the following steps:

1) Normalize the testing password feature as shown in (18) and obtain $\tilde{F}_{test}$. Calculate the Hamming distance $D_H$ between the given testing password and the password templates.

2) If $d_H$ is greater than the threshold $T_{D_H}$, reject the given testing password.

3) If $d_H$ is smaller than the threshold $T_{D_H}$, authenticate the given testing password. Let $\Delta=F_{test}-\hat{\mu}$ 

4) Update $\hat{\mu}_{new}=\hat{\mu}+\eta\Delta$, where $\eta$ is a tuning parameter that represents the adaptive rate.

By doing so, the template mean is shifted toward the feature that generated from the new authenticated password. This allows us to accommodate the user's password variation over time. We keep the standard deviation $\hat{\sigma}$ unchanged through the adaptive update process because it is obtained from the training set which is generated by the corresponding user within a short period of time and it represents the size of the user's password interval. Our assumption is that the user's password interval will not change over time and thus only the template mean is updated through the process.
\section{Results}
\subsection{Parameter Tuning}
For the user authentication, we first focus on tuning the parameter in both Euclidean distance based and Hamming distance based methods to achieve the best authentication performance. Therefore, a leave-one-out cross-validation is conducted for all 3 tasks (signatures with the static device, signatures with holding device and patterns with holding device) by using the training data only for both distance based methods. We use the authentication accuracy when the false match rate (FMR) is 0.1 \% as a metric to determine the parameter.

\begin{figure}[thpb]
	\includegraphics[width=0.48\columnwidth]{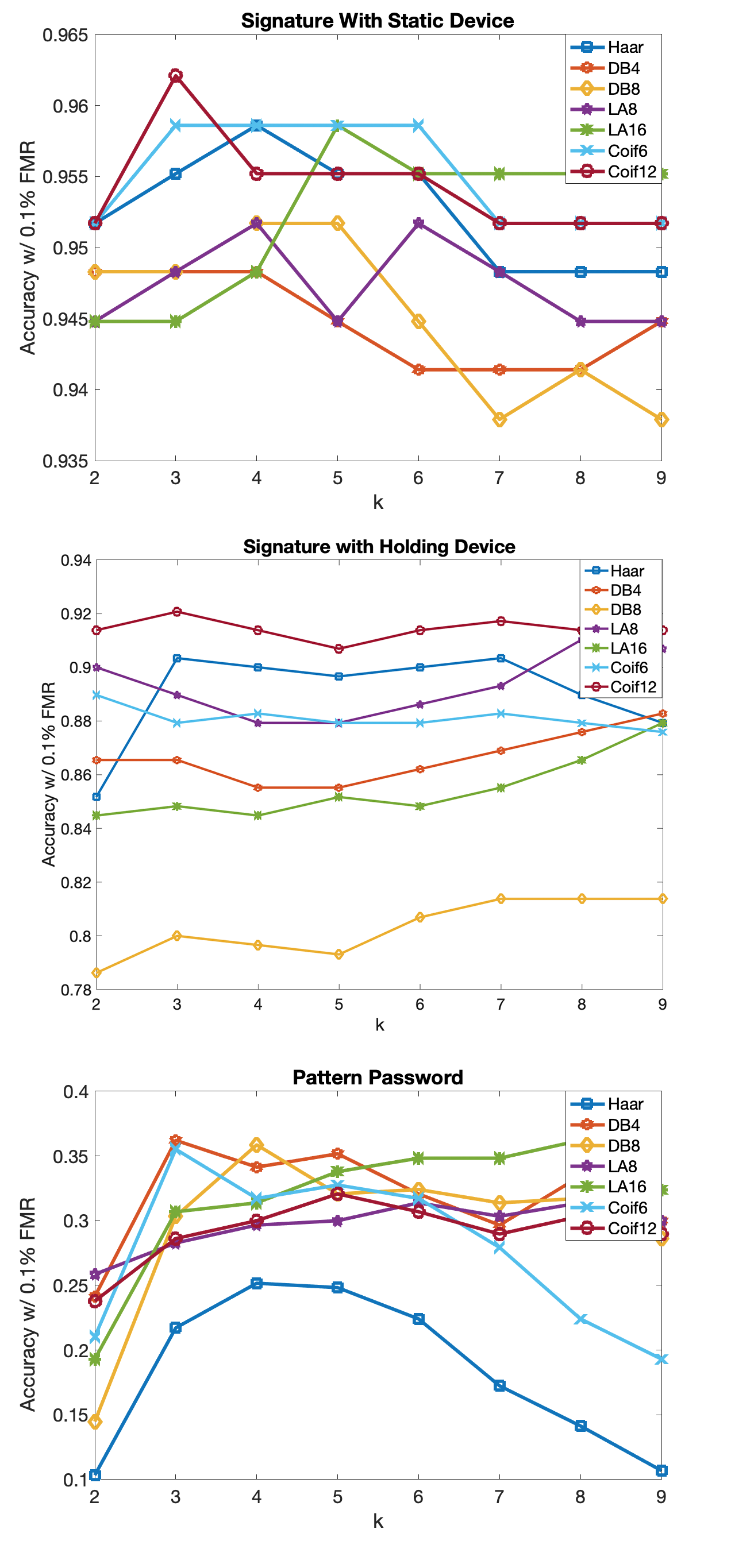}
	\includegraphics[width=0.48\columnwidth]{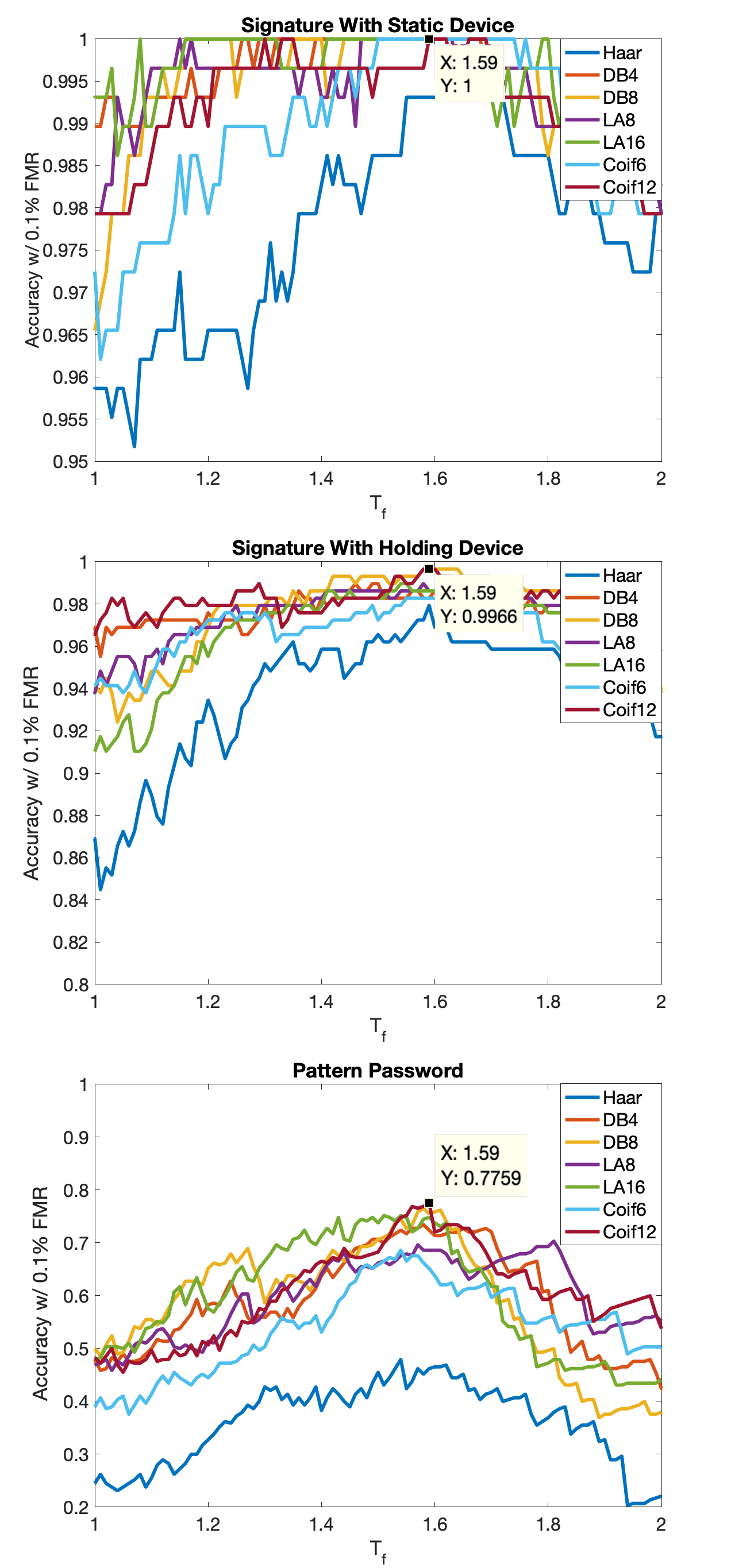}
	\caption{Euclidean distance based parameter tuning(left){\tiny } and Hamming distance based parameter tuning (right) Devices. For Euclidean distance based method, the best authentication performance is achieved when: 1) signature with static and holding device:  Coif12 as mother wavelet and $k=3$; 2) pattern: DB4 as mother wavelet and $k=3$ . For Hamming distance based method, the best authentication performance is achieved when we choose Coif12 as the mother wavelet and $T_f=1.59$ for all 3 tasks.}~\label{fig_res:figure1}
\end{figure}

For the Euclidean distance based method, we want to find the combination of the mother wavelet and the number of nearest template feature set $k$ (as shown in (17)) that achieves the best authentication performance. Therefore, we tested the Haar wavelet, Daubechie wavelet (DB4 and DB8), least asymmetric wavelet (LA8 and LA16) and coiflet wavelet (C6 and C12) as the mother wavelet. The number of nearest template feature set $k$ is varied from 2 to 9. The cross-validation results of the Euclidean distance based method are shown in Fig. \ref{fig_res:figure1} left.

We noticed that when we use Coif12 as the mother wavelet and $k=3$, the best authentication performance can be obtained for signature with the static and holding device. For the pattern-based password, the best authentication performance occurs when we use DB4 as the mother wavelet and $k=3$. Therefore, for the Euclidean distance based method, we choose Coif12 as mother wavelet and $k=3$ for both signatures with static and holding devices, and choose DB4 as the mother wavelet and $k=3$ for pattern-based passwords.

Similarly, for the Hamming distance base method, we want to find the combination of the mother wavelet and the feature difference threshold $T_f$ that achieves the best authentication performance. In addition to test aforementioned mother wavelets, $T_f$ is varied from 0 to 5 and the corresponding cross-validation results (around the peak) are shown in Fig. \ref{fig_res:figure1} right.

We noticed that for all 3 tasks, when we use Coif12 as mother wavelet and $T_f = 1.59$, the best authentication performance is achieved. Therefore, for the Hamming distance based method, we choose Coif12 as mother wavelet and $T_f=1.59$ for all 3 tasks.




\subsection{Authentication performance}
Authentication performance was evaluated using the genuine testing sets and the corresponding forgeries. Our evaluation metric includes the false match rate (FMR), the false non-match rate (FNMR), and the equal error rate (EER). FMR is the percentage of the forgeries from malicious individuals that are matched as genuine passwords by the authenticator. FNMR is the percentage of the genuine passwords from a legitimate user that are not matched as genuine passwords the authenticator. In most classification and authentication systems, there is a trade-off between FMR and FNMR. By adjusting the threshold so that the classifier is more strict and sensitive, one can reduce the FMR since fewer forgeries will be accepted by the system. However, FNMR will increase inevitably as more genuine passwords will be rejected due to the strict classifier. Therefore, in order to address this tradeoff, we use EER to represent authentication performance. The equal error rate is the rate at which both FMR and FNMR are equal.

\begin{figure}[thpb]
	\includegraphics[width=0.65\columnwidth]{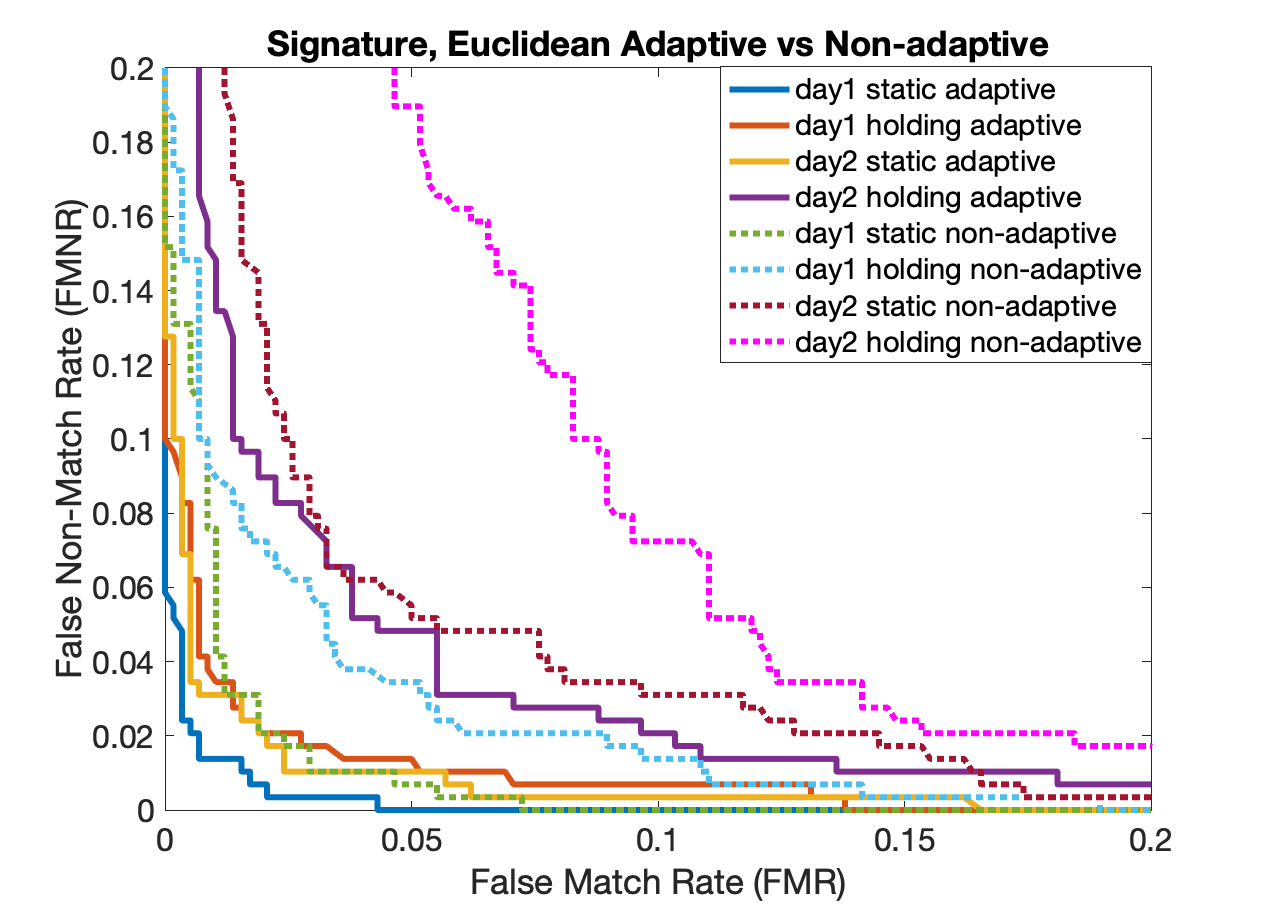}
	\includegraphics[width=0.65\columnwidth]{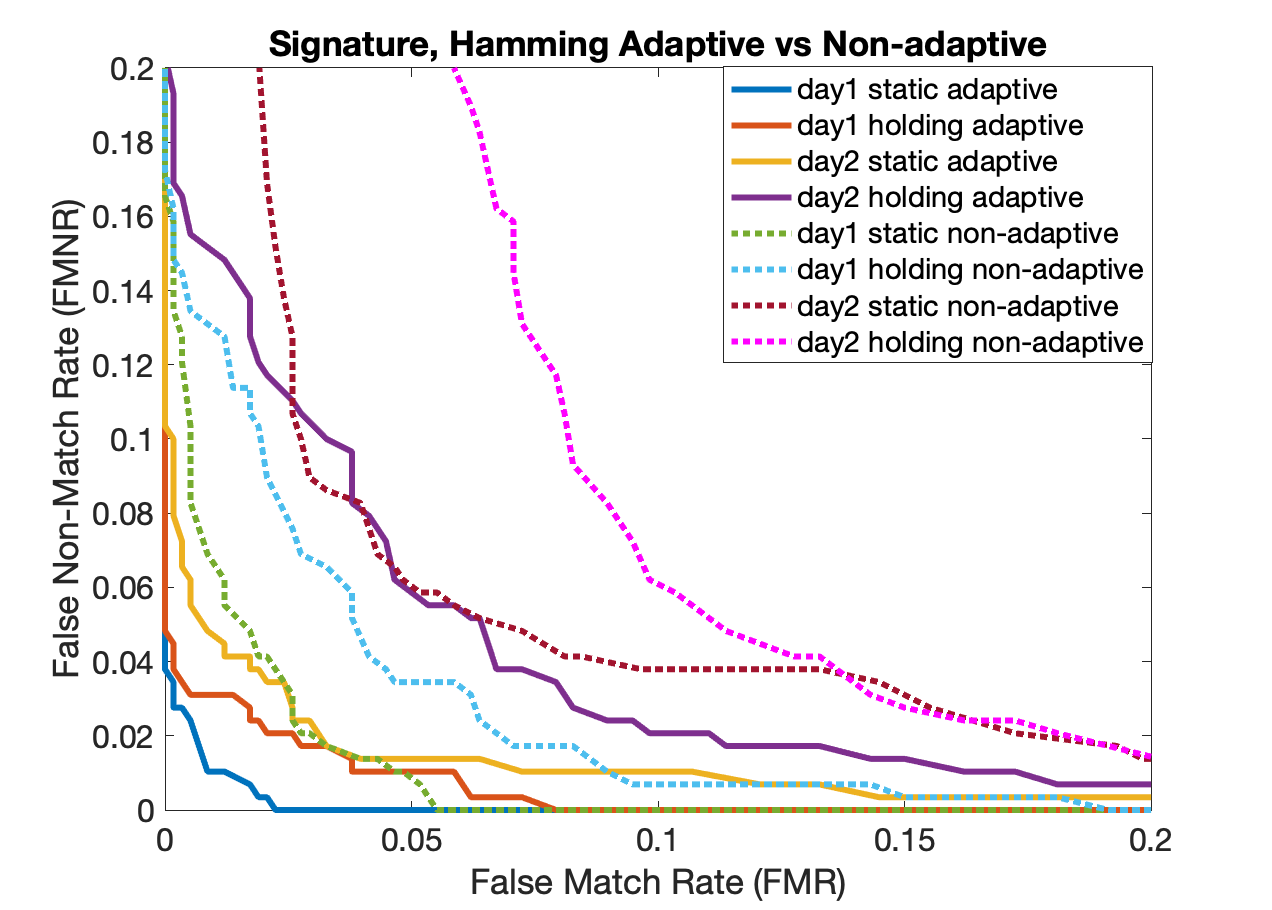}
	\caption{DET Curve for Signature Based Password, Euclidean Distance Based (top) and Hamming Distance Based (bottom) Method}~\label{fig_det:figure1}
\end{figure}

\begin{figure}[thpb]
	\includegraphics[width=0.65\columnwidth]{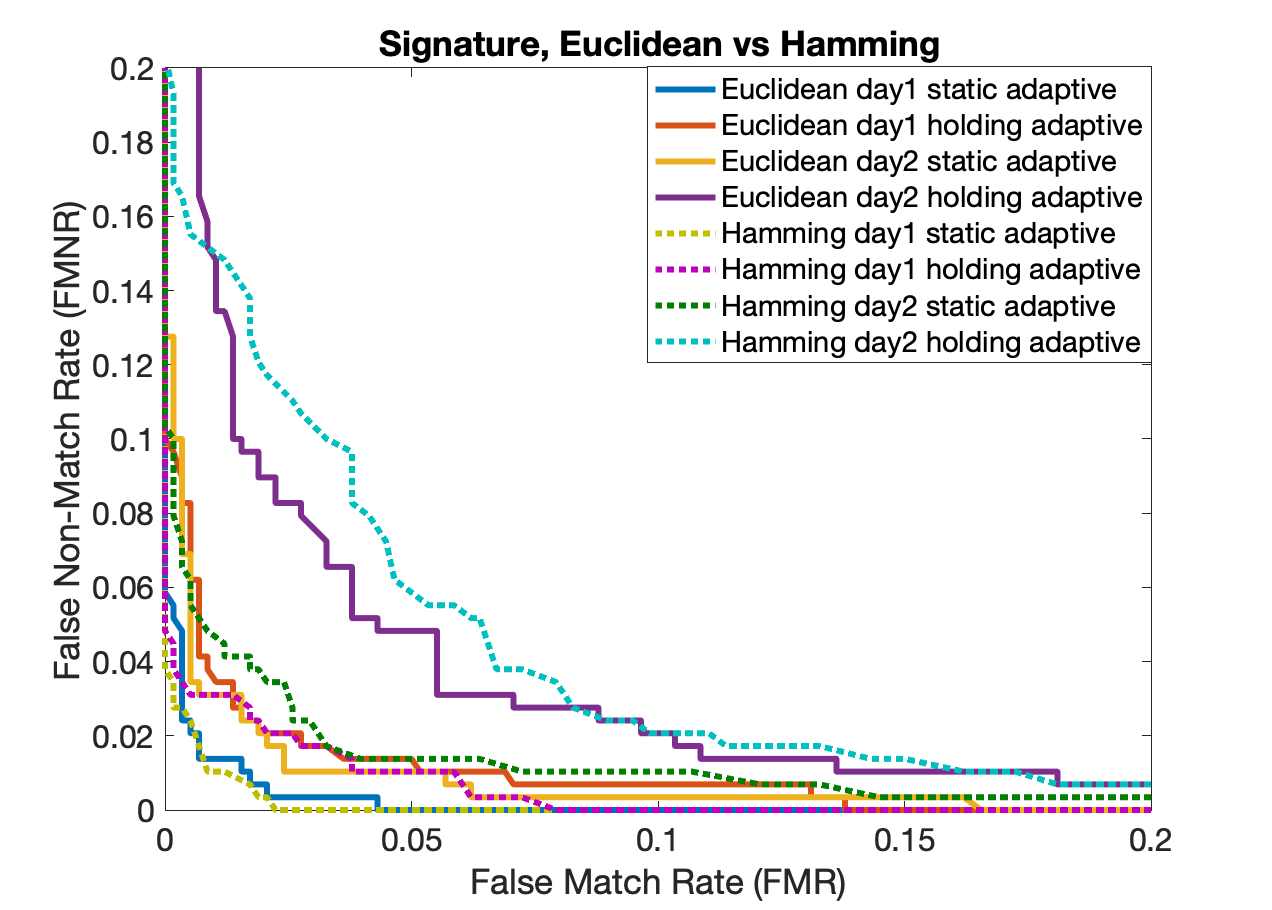}
	\caption{DET Curve for Signature Based Password, Euclidean Distance Based vs Hamming Distance Base Method}~\label{fig_det:figure2}
\end{figure}

\begin{figure}[thpb]	
	\includegraphics[width=0.65\columnwidth]{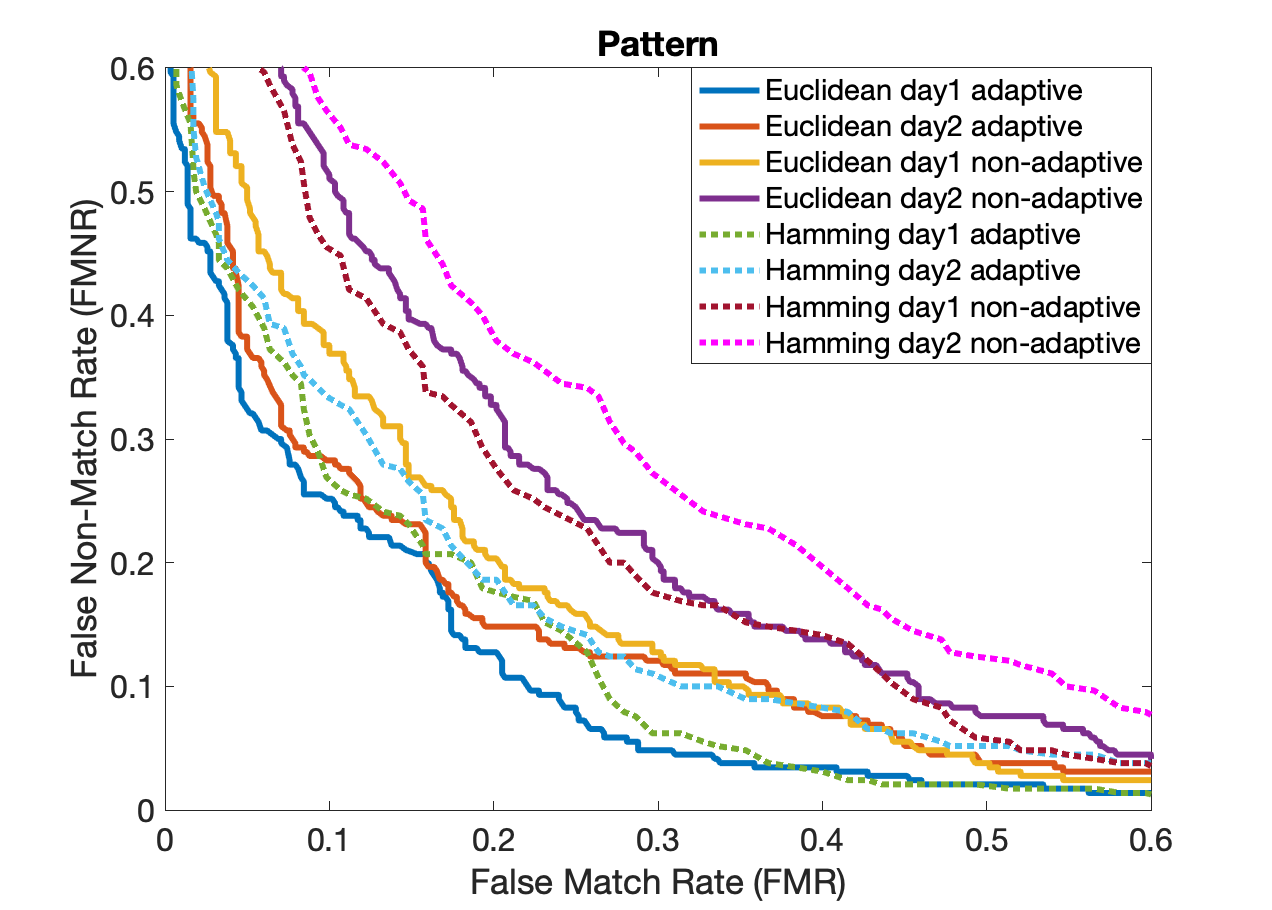}
	\caption{DET Curve for Pattern Based Password}~\label{fig_det:figure3}
\end{figure}

Fig. \ref{fig_det:figure1}-\ref{fig_det:figure2} shows the detection error tradeoff (DET) curve for Euclidean distance based and Hamming distance based method of the signature based password. Fig. \ref{fig_det:figure3} represents the detection error tradeoff (DET) curve for the pattern based password. 

For an authentication system, especially when dealing with life-critical systems, false matches are far more detrimental than false non-matches. Therefore, we also include the authentication accuracy when FMR is 0.1\% as a performance metric.
Tables \ref{tab:res_table2} and \ref{tab:res_table3} show the authentication performance for the Euclidean distance based and Hamming distance based methods.


\begin{table}
	\caption{Euclidean Distance Based Authentication Performance}
	\label{tab:res_table2}
	\begin{center}
		\begin{tabular}{c c c c c c}
			\hline
			\multirow{2}{*}{Task} & \multirow{2}{*}{Session Day} & \multicolumn{2}{c}{W/O Adaptive} & \multicolumn{2}{c} {W/ Adaptive}\\
			\hhline{~~----}
			& & EER & Accuracy w/ FMR = 0.1 & EER & Accuracy w/ FMR = 0.1\%  \\
			\hline
			Signature w/ static device & Day 1 & 2.07\% & 84.83\% & 1.38\% & 94.14\%\\
			Signature w/ static device & Day 2 & 5.17\% & 45.86\% & 2.07\% & 87.24\%\\
			Signature w/ holding device & Day 1 & 3.79\% & 81.03\% & 2.07\% & 90.00\%\\
			Signature w/ holding device & Day 2 & 8.97\% & 40.00\% & 4.83\% & 74.48\%\\
			Pattern w/ holding device & Day 1 & 20.52\% & 17.24\% & 16.90\% & 38.62\%\\
			Pattern w/ holding device & Day 2 & 24.14\% & 4.48\% & 17.24\% & 30.69\%\\
			\hline

		\end{tabular}
		\newline
		\newline
		\newline
		
	\end{center}
	\caption{Hamming Distance Based Authentication Performance}
	\label{tab:res_table3}
	\begin{center}
		\begin{tabular}{c c c c c c}
			\hline
			\multirow{2}{*}{Task} & \multirow{2}{*}{Session Day} & \multicolumn{2}{c}{W/O Adaptive} & \multicolumn{2}{c} {W/ Adaptive}\\
			\hhline{~~----}
			& & EER & Accuracy w/ FMR = 0.1 & EER & Accuracy w/ FMR = 0.1\%  \\
			\hline
			Signature w/ static device & Day 1 & 2.84\% & 83.79\% & 0.96\% & 95.86\%\\
			Signature w/ static device & Day 2 & 6.71\% & 60.34\% & 3.26\% & 88.97\%\\
			Signature w/ holding device & Day 1 & 4.47\% & 78.97\% & 2.33\% & 90.34\%\\
			Signature w/ holding device & Day 2 & 9.95\% & 43.79\% & 6.09\% & 77.24\%\\
			Pattern w/ holding device & Day 1 & 22.83\% & 17.93\% & 19.05\% & 32.76\%\\
			Pattern w/ holding device & Day 2 & 28.09\% & 8.28\% & 20.23\% & 28.97\%\\
			\hline

		\end{tabular}
	\end{center}
\end{table}

\subsubsection{Authentication Performance Over Multiple Days}
We see from Tables  \ref{tab:res_table2}-\ref{tab:res_table3}, and Fig. \ref{fig_det:figure1}, that, when no adaptation is applied to the authentication process, the authentication performance in the day 2 session degrades significantly, as compared to day 1 session for all cases. This indicates that human haptic passwords do vary over time and it is necessary to compensate for this. We therefore analyze the impact that our proposed adaptive template update schemes have on authentication performance.

The results shown in Fig. \ref{fig_det:figure1} and \ref{fig_det:figure2} indicate that adaptive template update schemes enhance the authentication performance for both Euclidean distance based and Hamming distance based methods, especially for the day 2 sessions. This is an evidence that the adaptation schemes for both methods can compensate for the user's password variation over time, thus enhancing the system authentication performance. 

\subsubsection{Authentication Performance of the Euclidean and Hamming Distance Based Methods}
According to Tables  \ref{tab:res_table2}-\ref{tab:res_table3} and Fig. \ref{fig_det:figure2}, there is no significant authentication performance difference between the Euclidean and Hamming distance based methods.

Since the MODWT-based feature extraction captures both frequency and localized (in time) information, a feature vector, which has a small number of dominant features with a large offset to the template, can still be a genuine one, as the user might do something inconsistently. When multiple dimensions of a testing feature have a \textbf{medium offset}, the testing feature is more likely generated from a forgery, \emph{since medium offset in multiple dimensions of features implies the global difference}.

Euclidean distance based method can better handle the cases where the user has multiple types of genuine signatures. However, it can not distinguish the cases between the forgery password with medium offsets in multiple dimensions, and the genuine password with large offsets in a small number of dominant dimensions. For both cases, it will generate a similar matching distance as the Euclidean distance between both types of features and the template features are similar.

The Hamming distance based method is unable to do well when there are multiple types of genuine signatures in the training set. On the other hand, however, the Hamming distance based method is less sensitive to a single or small number of dimensions that dominate the template features. The Hamming distance between a genuine password with a single or a small number of large offset and template features is \textbf{small}, while the Hamming distance between the forgery password with multiple medium offsets and template features is \textbf{large}. 

Therefore, the Euclidean and Hamming distance based method generate similar authentication performance.

\subsubsection{Authentication Performance of Different Postures and Tasks}
We observed that performance is slightly better when the device is held on a static surface, than when it is held in hand. This may be due to inconsistency in how the user holds the phone.

The authentication performance of the pattern-based password is not as good as the signature-based password. This is likely because the password complexity for the pattern-based password is much smaller than the signature-based password.

\subsection{Effect of password complexity}
Next we consider how the complexity of a password affects authentication performance. Similar to alphanumerical passwords, we anticipate that the complexity of the user's haptic password will affect the authentication performance. To quantitatively evaluate the effect of the password complexity on the performance of the haptic passwords system, the signature complexity $\delta$ of the user $i$ is defined as
$$
\delta_i=\frac{s_i}{s_{max}}+\frac{x_i}{x_{max}}+\frac{a_i}{a_{max}}+\frac{d_i}{d_{max}}\eqno{(19)}
$$
\\
Here \\
1) $s$ represents the average number of strokes in the given user's password in the training set. Within each stroke, the user's finger is continuously contacting the phone screen. Two consecutive strokes are separated by a period when there is no contact between the user's finger and the phone screen;\\
2) $x$ represents the average number of trajectory intersections in the training set;\\
3) $a$ represents the average number of acceleration zero crossings in the training set;\\
4) $d$ represents the portion of the curved trajectory in the password. We add $d$ to the signature complexity because many subjects in our experiments expressed that it is more difficult to mimic curved lines, compared to straight lines and vertices (i.e. sharp turnings). The portion of the curved trajectory in the signature is calculated by averaging the number of medium trajectory direction changes in the training set. The medium trajectory direction change is defined as $|\omega(t)|\in[\pi/6,\pi/4]$;\\

$[\cdot]_i$ represents the corresponding value of user $i$ and $[\cdot]_{max}$ represents the largest value across all users within the same task.

The password complexity $\delta\in[0,4]$, as all 4 metrics listed above are rescaled to $[0,1]$ based on the maximum value in the corresponding metric among all subjects. The larger the complexity is,  the more complex the user's signature will be.

\begin{figure}
	\includegraphics[width=0.65\columnwidth]{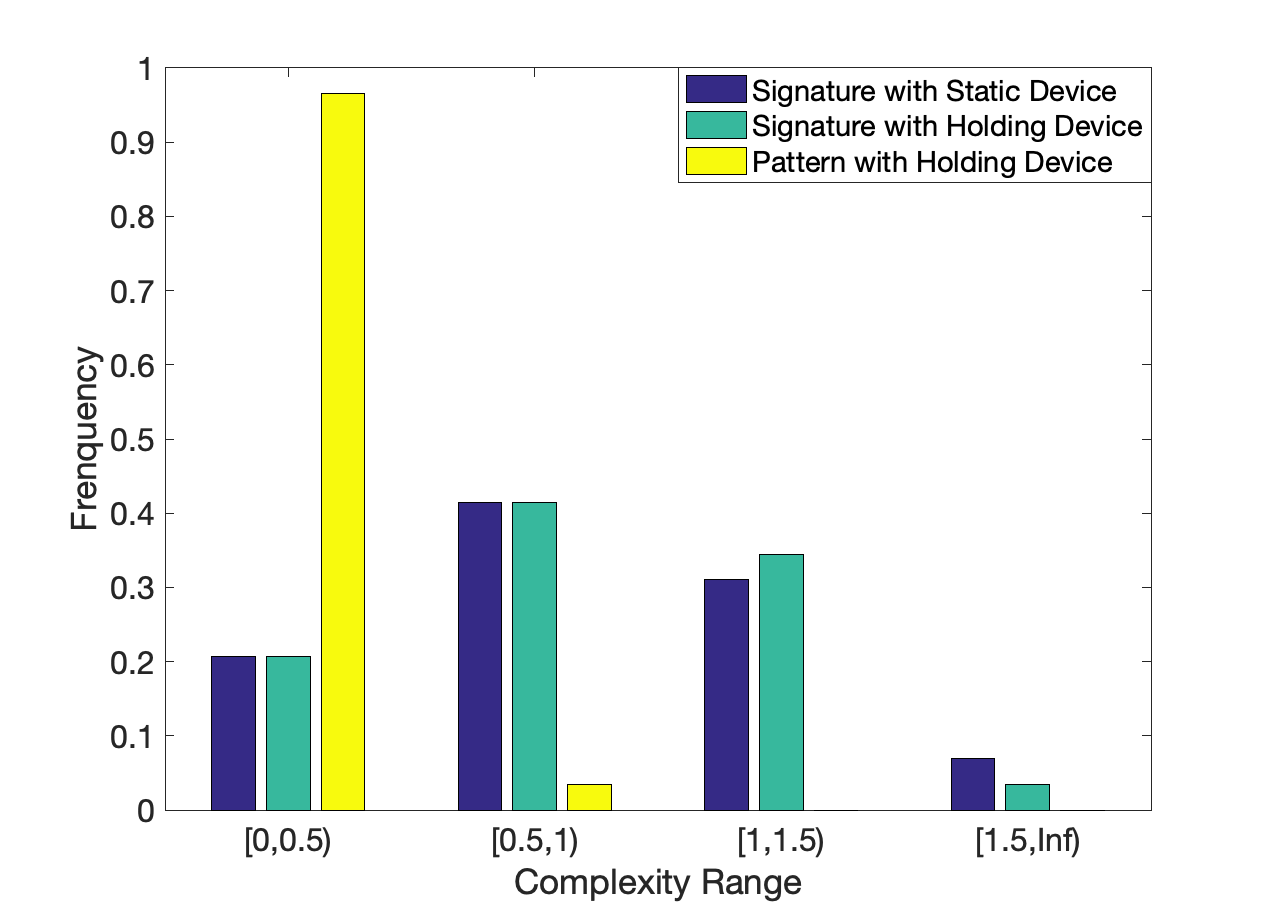}
	\caption{Complexity Histogram}~\label{fig_res:figure8}
\end{figure}

Fig. \ref{fig_res:figure8} shows the complexity histogram of signatures when the device is on a static surface and when it is held. The complexity of almost all pattern passwords is within the low complexity range [0,0.5) while the signature-based passwords are more complex. 

Intuitively, a simple password is more likely to be forged. To prove this intuition, we first define the forging difficulty of each user's password as the smallest Hamming distance any forgery can achieve among all 20 forgeries in the corresponding task. The smaller the forging difficulty is, the more likely the corresponding password can be forged.

\begin{table}
	\caption{Average Signature Forging Difficulty within Each Complexity Range}
	\label{tab:res_table4}
	\begin{center}
		\begin{tabular}{c c c c c }
			\hline
			\multirow{2}{*}{Task}  & \multicolumn{4}{c}{Complexity Range}\\
			\hhline{~----}
			& [0,0.5) & [0.5,1) & [1,1.5) & [1.5,$\infty$)  \\
			\hline
			Static Signature & 111.50 & 119.50 & 120.11 & 136.00\\
			Holding Signature & 105.67 & 114.50 & 121.00 & 129.50\\
			Holding Pattern  &107.32 & 105.00 & N/A & N/A\\
			\hline

		\end{tabular}
	\end{center}
\end{table}	

Table \ref{tab:res_table4} illustrate how the password complexity affects the corresponding forging difficulty. 
First, we notice that the forging difficulty of the pattern-based password is small, as all pattern-based passwords have a relatively small password complexity. This makes it easier to forge pattern-based passwords and the corresponding authentication performance for this task is not as good as the signature-based password. 
For both signatures with static and holding devices, the forging difficulty increases when the password complexity increases. This indicates that to achieve good authentication performance, it is crucial to guarantee that the original passwords in the training set are complex enough. A lack of complexity will increase the probability of the given password being forged. In practical applications, to deal with this issue, if a person's signature is too simple, they could also include writing some numbers, since people do that consistently, to increase the password complexity.

\subsection{Password Entropy}
The possible password space and scalability are crucial for the password system application. In this section, we investigate the password entropy of the haptic password system as an indicator of the system scalability. For a password of Length $L$, the definition of password entropy\cite{florencio2007large} is shown in (20), where $S_i$ is the number of possible symbols at $i^{th}$ location.

$$
H = \sum_{i=1}^{L} log_2 S_i\eqno{(20)}
$$

The haptic password entropy is calculated in the following steps.

1. \textbf{Find uncorrelated features to calculate the password entropy.} Since velocity, acceleration, trajectory direction and trajectory direction change as shown in (1) - (5) are generated from the position, we only consider the position and force features. We then calculate pairwise correlation among the selected features and discard those features with an average correlation above 0.3 to obtain uncorrelated features.

2. \textbf{Calculate possible symbol size for each uncorrelated feature.} First, for each feature $f_i$, we find the value range $R_i$ among all users as (21).

$$
R_i = f_{i,max}-f_{i,min}\eqno{(21)}
$$

We then calculate the standard deviation $\sigma_i$ for each feature $f_i$ for each subject and obtain the largest standard deviation among all the subject as $\sigma_{i,max}$. The possible symbol size $S_i$ for feature $f_i$ is defined as in (22)

$$
S_i = \frac{R_i}{3\sigma_{i,max}}\eqno{(22)}
$$

3. \textbf{Calculate Password Entropy.} Given the possible symbol size of each feature, the password entropy can be calculated as shown in (20).

By following the above steps, we obtain the haptic password entropy for signature-based passwords is 67.45 bits. The average entropy of an alphanumerical password is 40.54 bits\cite{florencio2007large}.

\section{Open Questions}
\textbf{Cross posture authentication.} As investigated in the previous section, the authentication performance of the signature-based password is good within static posture and holding posture. An interesting question is can we generate a unified template that can be used to authenticate users in both postures. However, our analysis shows that when we use the signature with the static device template to authenticate the testing signature with the holding device (and vice versa), we are unable to achieve as good an authentication performance as within each posture. The reason for such an outcome is two-fold. First, when holding the mobile phone in the hand, the mobile phone position is less stable which makes the signature trajectory deviate from those static signatures. Second, the force applied to the touch screen is different in each posture. For the signature with the static device, the phone is put on a rigid surface while for the signature with the holding device, the phone is held by the user in hand. The user's hand can be regarded as an elastic surface. This difference will affect the force applied to the screen as the holding posture will work as a low-pass filter and filter out local force details. Moreover, given the difference between static and holding posture, the proposed adaptive template update scheme might not fit. Nevertheless, however, there is a possible way to overcome this difficulty. With the gyrometer and accelerometer data on the mobile phone, it is relatively easy to determine the posture of the user and the user can be authenticated using the corresponding template based on the posture determined.

\textbf{Long term adaptive template update.}
In the proposed experiment study, we were only able to conduct a 2-day experiment. We showed that the adaptive template update compensates users' variation over time on the second day. However, we notice the degradation across days still exists. How users' passwords vary over long-term and can the proposed password adaptive template update scheme accommodate the variation is worth further investigation.

\textbf{Templates poisoning caused by adaptive template update.}
Although the adaptive template update method enhances the system authentication accuracy, if not used properly, attackers can take advantage of this mechanism and progressively change the genuine user's template set to the attack's template set. There are two methods to mitigate this issue:

1) When determining the threshold, choose the one that generates low FMR (strict threshold). In this way, it will be harder for an attacker to generate a forgery to be authenticated. Although, this will increase the false non-match rate, according to Table \ref{tab:res_table2}-\ref{tab:res_table3}, with 0.1\% FMR, the probability, which a genuine user has to re-enter more than 3 times to be authenticated, is less than 1\%.

2) After every time the template is updated, calculate the difference between the current template and the original template obtained during initial training. For example, for the Hamming distance based method, calculate the distance between the current "center" and the initial training "center". If the distance is larger than a certain threshold, which means the current template has been changed significantly, request the user to provide additional authentication and retrain the model with a new training session.

The first method reduces the possibility that an attacker injects a forgery to the template. Even if any forgery passed the authentication, the second method prevents the attacker from progressively taking over the template.

\textbf{Different forms of the password.} A possible extension of our work is to investigate other forms of passwords besides signatures and simple patterns. Every user potentially has a unique way of handwriting. Therefore, as long as the person does some writing tasks, such as a word, a series of digits or even a doodle of a cat, frequently so that they are consistent, they can be a password. Extra experiments need to be conducted to examine the password authentication performance of various forms.
\section{Conclusion}
In this paper, a haptic password system on a mobile device platform was developed. We investigated whether and how the way a user haptically interacts with a force-sensing touchscreen can serve as a behavioral biometric for user authentication. We designed and conducted a human subject experimental study to collect both signature-based and pattern-based passwords of 29 subjects. We developed a MODWT feature extraction along with customized classifiers to fulfill the user authentication process. To accommodate users' password variation over time, we also proposed an adaptive template update scheme. The authentication process is able to achieve robust authentication results for the signature-based password, with equal error rates 0\%-5\% and 75\%-96\% accuracy with 0.1\% FMR under forgery attacks depending on the experiment setting. Moreover, we investigated the effect of password complexity on the authentication performance and quantitatively demonstrated that
the more complex a password is, the less likely it can be forged. The pattern-based passwords are unable to achieve as good performance as the signature-based password is because the complexity of the pattern-based password is small which makes it easy to be forged. We also showed the good scalability of the haptic passwords as the password entropy is much larger than alphanumerical passwords. The results indicate that the proposed method is able to generate robust forgery-proof authentication performance for user authentication on the force-sensing mobile devices.

Future work might involve analysis of how users' haptic passwords vary over the long term and exploration of the authentication performance of different forms of haptic passwords.

\begin{acks}
This work is supported by the National Science Foundation, Grant No. CNS-1329751, UW CoMotion, and the Amazon Catalyst program.
The authors would like to thank all BioRobotics Lab members for their help and insights on this work.
\end{acks}

\bibliographystyle{ACM-Reference-Format}
\bibliography{sample}
\end{document}